\documentclass{fundam}
\usepackage{graphicx}

\begin{document}
\setcounter{page}{1}
\issue{XXI~(2008)}

\title{Towards a stable definition of Kolmogorov-Chaitin complexity}

\author{Jean-Paul Delahaye\\
Laboratoire d'Informatique Fondamentale de Lille (CNRS)\\
jean-paul.delahaye{@}lifl.fr
\and Hector Zenil\thanks{Some of the ideas contained in this paper were developed during the stay of H. Zenil's tenure as a visiting scholar at Carnegie Mellon University. He wishes to thank Jeremy Avigad for his support and Kevin Kelly for his valuable comments and suggestions.}\\
Laboratoire d'Informatique Fondamentale de Lille (CNRS)\\
Institut d'Histoire et de Philosophie des Sciences et des Techniques, (CNRS, ENS, Universit\'e Paris 1)\\
Carnegie Mellon University\\
hector.zenil@lifl.fr, hector.zenil-chavez@malix.univ-paris1.fr, hectorz@andrew.cmu.edu} \maketitle

\runninghead{JP. Delahaye, H. Zenil}{Towards a stable definition of Kolmogorov-Chaitin complexity}

\begin{abstract}
Although information content is invariant up to an additive constant, the range of possible additive constants applicable to programming languages is so large that in practice it plays a major role in the actual evaluation of $K(s)$, the Kolmogorov-Chaitin complexity of a string $s$. Some attempts have been made to arrive at a framework stable enough for a concrete definition of $K$, independent of any constant under a programming language, by appealing to the \emph{naturalness} of the language in question. The aim of this paper is to present an approach to overcome the problem by looking at a set of models of computation converging in output probability distribution such that that \emph{naturalness} can be inferred, thereby providing a framework for a stable definition of $K$ under the set of convergent models of computation. \\
\end{abstract}


\section{Introduction}

We will use the term \emph{model of computation} to refer both to a Turing-complete programming language and to a specific device such a universal Turing machine. 

The term \emph{natural} for a Turing machine or a programming language has been used within several contexts and with a wide range of meanings. Many of these meanings are related to the expressive semantic framework of a model of computation. Others refer to how well a model fits with an algorithm implementation. Previous attempts have been made to arrive at a model of computation stable enough to define the Kolmogorov-Chaitin complexity of a string independent of the choice of programming language. These attempts have used, for instance, lambda calculus and combinatory logic\cite{li,tromp} appealing to their \emph{naturalness}. We provide further tools for  determining whether approaches such as these are natural to produce the same relative Kolmogorov-Chaitin measures. Our approach is an attempt to make precise such appeals to the term \emph{natural} related to the Kolmogorov-Chaitin complexity, and to provide a framework for a stable definition of $K$  independent enough of additive constants.\\

\texttt{Definition} The Kolmogorov-Chaitin  complexity $K_u(s)$ of a string $s$ with respect to a universal Turing machine $U$ is defined as the binary length of the shortest program $p$ that produces as output the string $s$.
\begin{center}
$K_u(s) = \{min(|p|), U(p)=s\}$
\end{center}

A major drawback of $K$ is that it is uncomputable\cite{calude} because of the undecidability of the halting problem. Hence the only way to approach $K$ is by compressibility methods. A major criticism brought forward against $K$ (for example in\cite{kelly}) is its high dependence of the choice of programming language.

\section{Dependability on additive constants}

The following theorem tells us that the definition of Kolmogorov-Chaitin complexity makes sense even when it is dependent upon the programming language:\\

\texttt{Theorem (invariance)}  If $L_1$ and $L_2$ are two Turing machines and $K_{L_1}(s)$ and $K_{L_2}(s)$ the Kolmogorov - Chaitin complexity of a binary string $s$ when $L_1$ or $L_2$ are used respectively, then there exists a constant $C_{L_1,L_2}$  such that for all binary string $s$:  

\begin{center}
$| K_{L_1}(s) - K_{L_2}(s) | < C_{L_1,L_2}$\\
\end{center}

In other terms, there is a program $p_1$ for the universal machine $L_1$ that allows $L_1$ to simulate $L_2$. This $p_1$ is usually called an interpreter or compiler in $L_1$ for $L_2$. Let $p_2$ be the shortest program producing some string $s$ according to $L_2$. Then the result of chaining together the programs $p_1$ and $p_2$ generates $s$ in $L_1$. Chaining $p_{L_2}$ onto $p_1$ adds only constant length to $p_2$, so there exists a constant $C$ that bounds the difference in length of the shortest program in $L_1$ from the length of the shortest program in $L_2$ that generates the arbitrary string $s$. 

However, the constants involved can be arbitrarily large so that one can even affect the relative order relation of  $K$ under two different universal Turing machines such that if $s_1$ and $s_2$ are two different strings and $K(s_1)<K(s_2)$ one can construct an alternative universal machine that not only changes the values for $K(s_1)$ and $K(s_2)$ but reverses the relation order to $K(s_1)>K(s_2)$.

One of the first conclusions drawn from algorithmic information theory is that at least one among the $2^n$ binary strings of length $n$ will not be compressible at all. That is because there are only $2^n-1$ binary programs shorter than $2^n$. In general, if one wants to come up with an ultimate compressor one can compress the length of every string by $c$ bits with $2^{n-c}$ length descriptions. It is  a straightforward conclusion that  no compressing language can arbitrarily compress all strings at once. The strings a language can compress depend  on the  language used, since any string (even a random-looking one) can in some way be encoded to shorten its description within the language in question even if a string compressible under other languages turns out to be incompressible under the new one. So one can always come up with another language capable of effectively compressing any given string. In other terms, the value of $K(s)$ for a fixed $s$ can be arbitrarily made up by constructing a suitable programming language for it. However, one would wish to avoid such artificial constructions by finding distinguished programming languages which are \emph{natural} in some technical sense--rather than tailor-made to favor any particular string-- while also preserving the relative values of $K$ for all (or most) $2^n$ binary strings of length $n$ within any programming language sharing the same order-preserving property. 

As suggested in \cite{kelly}, suppose that in a programming language $L_1$, the shortest program $p$ that generates a random-looking string $s$ is almost as long as $s$ itself. One can specify a new programming language $L_2$ whose universal machine $U_2$ is just like the universal machine $U_1$ for $L_1$ except that, when presented with a very short program $p_2$ , $U_2$ simulates $U_1$ on the long program $p$, generating $s.$ In other words, the complexity of $p$ can be "buried" inside of $U_2$ so that it does not show up in the $U_2$ program $p_2$ that generates $s$. This arbitrariness makes it hard to find a stable definition of Kolmogorov-Chaitin  complexity unless a theory of \emph{natural} programming languages is provided which is unlike the usual definition in terms of an arbitrary, Turing-complete programming language.

For instance, one can conceive of a universal machine that produces certain strings very often or very seldom, despite being able to produce any conceivable string given its universality. Let's say that a universal Turing machine is tailor-made to produce much fewer $(0)^n$ strings than any other string in $s \in \{0,1\}^n$. By following the relation of Kolmogorov-Chaitin  complexity to the universal distribution\cite{solomonoff,levin} $m(s)=1/2^{K(s)+O(1)}$ one would conclude that for the said tailor-made construction the string $(0)^n$ is of greater Kolmogorov-Chaitin  complexity than any other, which may seem counterintuitive. This is the kind of artificial constructions one would prefer to avoid, particularly if there is a set of programming languages for which their output distributions converge, such that between two \emph{natural} programming languages the additive constant remains small enough to make $K$ invariant under the encoding from one language to the other, thus yielding stable values of $K$. 

The issue of dependence on additive constants often comes up when $K$ is evaluated using a particular programming language or universal Turing machine. One will always find that the additive constant is large enough to produce very different values. This is even worst for short strings, shorter for instance compared to the program implementation size. One way to overcome the problem of the calculation of $K$ for short strings was suggested in \cite{delahayezenil1,delahayezenil2}. It involved building from scratch a prior empirical distribution of the frequency of the outputs according to a formalism of universal computation. In these experiments, some of the models of computation explored seemed to converge, up to a certain degree, leading to propose a natural definition of $K$ for short strings. That was possible because the additive constant up to which the output probability distributions converge has a lesser impact on the calculation of $K$, particularly for those at the top of the classification (thus the most frequent and stable strings). This would make it possible to establish a stable definition and calculation of $K$ for a set of models of computation identified as natural  for which $K(s)$ relative orders are preserved even for larger strings. 

Our attempt differs from previous attempts in that the programs generated by different models may produce the same relative $K$ despite the programming language or the universal Turing machine being necessarily compact in terms of size. This is what one would expect for a stable definition of $K$ to work with even if there were still some additive constants involved.

\section{Towards a stable definition of $K$}
\label{experiment}

The experiment described in detail in  \cite{delahayezenil1} proceeded by analyzing the outputs of two different models of computation: deterministic Turing machines ($TM$) and one-dimensional cellular automata ($CA$). Some followed methods and techniques for enumerating, generating and performing exhaustive searches are suggested in further detail in \cite{wolfram}. The Turing machine ($TM$) model, represents the basic framework underlying many concepts in computer science, including the definition of Kolmogorov-Chaitin complexity, while cellular automaton, has been largely studied as a particular interesting model also capable of universal computation. The descriptions for both $TM$ and $CA$ followed standard formalisms commonly used in the literature. The Turing machine description consisted of a list of rules (a finite program) capable of manipulating a linear list of cells, called the \emph{tape}, using an access pointer called the \emph{head}. The directions of the tape are designated \emph{right} and \emph{left}. The finite program can be in any one of a finite set of states $Q$ numbered from 1 to $n$ with 1 the state at which the machine starts its computation. There is a distinguished $n+1$ state called the halting state at which the machine halts. Each tape cell can contain a 0 or 1 (there is no special blank character). Time is discrete and the time instants (steps) are ordered from $0,1,\ldots$ with 0 the time at which the machine starts its computation. At any time, the head is positioned over a particular cell. At time 0 the head is situated on a distinguished cell on the tape called the start cell, and the finite program starts in the state 1. At time 0 all cells contain the same symbol, either 0 or 1. A rule can be written in a $5$-tuple notation as follows $\{s_i,k_i,s_{i+1},k_{i+1},d\}$, where $s_i$ is the scanned symbol under the head, $k_i$ the state at time $t$, $s_{i+1}$ the symbol to write at time $t+1$, $k_{i+1}$ and $d$ the head movement either to the right or to the left at time $t+1$. As usual a Turing machine can perform the following operations:  1. write an element from $A=\{0,1\}$. 2. shift the head one cell left or right. 3. change the state of the finite program out of $Q$. And when the machine is running it executes the above operations at the rate of one operation per step. At the end of a computation the Turing machine has produced an output described by the contiguous cells in the tape over which the head went through.

An analogous standard description of a one-dimensional cellular automata was followed. A one-dimensional cellular automaton is a collection of cells on a grid that evolves through a number of discrete time steps according to a set of rules based on the states of neighboring cells that are applied in parallel to each row over time. In a binary cellular automaton, each cell can take only one among two possible values (0 or 1). When the cellular automaton starts its computation, it applies the rules at row 0. A neighborhood of $m$ cells means that the rule takes into consideration the value of the cell itself, $m$ cells to the right and $m$ cells to the left in order to determine the value of the next cell at row $n+1$.

For the Turing machines the experiments were performed over the set of 2-state 2-symbol Turing machines, henceforth denoted as $TM(2,2)$. There are $4096$ different Turing machines according to the description given above and the derived formula $(2sk)^{sk}$ from the traditional $5$-tuplet rule description of a Turing machine. It was then let all the machines run for $t$ steps each and proceeded to feed each with an empty tape with 0 and once again with an empty tape filled with 1.

It was proceeded in the same fashion for cellular automata with nearest-neighbor taking a single $1$ on a background of $0$s and a single start cell $0$ on a background of $1$s, henceforth denoted by $CA(1)$. There are $2\times2\times2=2^3=8$  possible binary states for the three cells neighboring a given cell, there are a total of $2^8=256$ elementary cellular automata or $ECA$.

Let $s(TM(i), m)$ and $s(CA(j), m)$ be the two sets of output strings produced by the $i$-th Turing machine and the $j$-th cellular automaton respectively, after $m$ steps according to an enumeration for Turing machines and cellular automata, a probability distribution was built as follows: the sample space associated with the experiment is $S = \{s | s \in \{0,1\}^n\}$ since both $s(TM(i), m)$ and $s(CA(j), m)$ are sets of binary strings. Let's call $S$ the set of outputs either from $s(TM(n), m)$ or $s(CA(n), m)$. For each $s \in S$ the space of the random variable $X \in S$ is $\{0,1\}^n$. For a discrete variable $X$, the probability $Pr(X=s)$ means the probability $f(x)$ of the random variable $X$ to produce the substring $s$. Let $D(X)=\{s_t, f(s_t)\}$ such that for all $s_i \in S$, $f(s_t)>f(s_{t+1})$. $f(x)$ is the probability of $s$ to be produced. In other words, $D(X)$ is the set of tuples of a string followed by the probability of that that string to be produced by a Turing machine or a cellular automata after $m=10n$ steps.

\subsection{Output probability distribution D(X)} 

$D(X)$ is a discrete probability distribution since $\Sigma_u Pr( X = s ) = 1$, as $u$ runs through the set of all possible values of $X$, for a set of finite number of possible binary strings, and the sum of all of them is exactly 1. $D(X)$ simply denoted as $D$ from now on was calculated in \cite{delahayezenil1} for two sets of Turing machines and cellular automata with small  state and symbol values up to certain string length $n$.

\begin{figure}
\begin{center}
\includegraphics[width=0.5\textwidth]{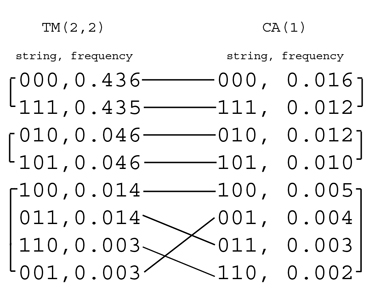}
\end{center}
\caption{The experiments can be summarized by looking at the above diagram comparing two output probability distributions for strings of length $n = 3$, after $t = n \times 10 = 30$ steps. Matching strings are linked by a line. As one can observe, in spite of certain crossings, $TM(2,2)$ and $CA(1)$ seem to be strongly correlated and both group the output strings by reversion and complementation symmetries. By taking the six groups--marked with brackets--both probability distributions make a perfect match. In \cite{delahayezenil1} we provide another example for strings length $n = 4$.}
\label{comparison}
\end{figure}

In each case $D$ was found to be stable under several variations such as number of steps and sample sizes, allowing to define a stable distribution $D$ for each, denoted from now on as $D_{TM}$ for the distribution of Turing machines and $D_{CA}$ for the distribution from cellular automata.\\

\subsection{Equivalence of complexity classes}

The application of a widely used theorem in group theory may provide further stability, getting rid of crossings due to exchanged strings, with different strings probably having the same Kolmogorov-Chaitin complexity but biasing the rank comparisons. Desirably, one would have to group and weight the frequency of the strings with the same expected complexity in order to measure the rank correlation without any additional bias. Consider, for instance, two typical distributions $D_1$ and $D_2$ for which the calculated frequency have placed the strings $(0)^n$ and $(1)^n$ at the top of $D_1$ and $D_2$ respectively. If the ranking distance of both distributions is then calculated, one might get a biased measurement due to the exchange of  $(0)^n$ with $(1)^n$ despite the fact that both should have, in principle, the same Kolmogorov-Chaitin complexity. Therefore, we want to find out how to group these strings such that after comparison they do not affect the rank comparison.

The P\'olya-Burnside enumeration theorem\cite{redfield} makes possible to count the number of discrete combinatorial objects of a given type as a function of their symmetrical cases was used. We have found that experimentally symmetries that are supposed to preserve the Kolmogorov-Chaitin complexity of a string are reversion $(re)$, complementation $(co)$ and the compositions from them ($cosy(s)$ and $syco(s)$). In all the distributions built from the experiments so far we have found that strings always tend to group themselves in contiguous groups with their complemented and reversed versions. That is also a consequence of the setting up of the experiments since each Turing machine ran from an empty tape filled with zeros first and then again with an empty tape filled with ones in order to avoid any antisymmetry bias. Each cellular automata ran starting with a 0 in a background of ones and once again with a 1 in a background of zeros as well for the same reason.\\

\texttt{Definition (complexity class)} Let $D$ be the probability distribution produced by a computation. A complexity class $C$ in $D$ is the set of strings \{$s_1$,$s_2$,\dots,$s_i$\} such that $K(s_1) = K(s_2) = \ldots = K(s_i)$.\\

The above clearly induces a partition since $\bigcup_{i=1}^n C_i=D$ and $\bigcap_{i=1}^n C_i=\emptyset$ for $n$ the number of strings in $D$. In other words, all strings in $D$ are in one and only one complexity class. We will denote $D_r$ the reduced distribution of $D$. Evidently the number of elements in $D$ is greater than or equal to $D_r$.

The P\'olya-Burnside enumeration theorem will help us arrive at $D_r$. There are $2^n$ different binary strings of length $n$ and 4 possible transformations to take into consideration:

\begin{enumerate}
\item  $id$, the identity symmetry, $id(s)=s$.
\item $sy$, the reversion symmetry given by: If $s=d_1d_2,\ldots d_n$, $sy(s)=d_nd_2,\ldots d_1$.
\item $co$, the complementation symmetry given by $co(s)=mod(d_i+1, 2)$.
\end{enumerate}

Let $T$ denote the set of all possible transformations under composition of the above.\\

\noindent The classes of complexity can then be obtained by applying the Burnside theorem according to the following formula: 

\begin{center}
$(2^n + 2^{n/2} + 2^{n/2})/4$, for $n$ odd\\
$(2^n + 2^{(n + 1)/2})/4$ otherwise.
\end{center}

This is obtained by calculating the number of invariant binary strings under $T$. For the transformation $id$ there are $2^n$ invariant strings. For $sy$ there are $2^{n/2}$ if $n$ is even, $2^{(n + 1)/2}$ if $n$ is odd, the number of invariant strings under $co$ is zero and the number of invariant strings under $syco$ is $2^{n/2}$ if $n$ is even, or zero if it is odd. Let's use $B(D)$ to denote the application of the Burnside theorem to a distribution $D$. As a consequence of applying $B(D)$, grouping and adding up the frequencies of the strings, once has to divide the frequency results by $2$ or $4$ (depending on the number of strings grouped for each class) according to the following formula:

\begin{center}
$f_r(s)/|\bigcup{(sy(s), co(s), syco(s))}|$
\end{center}
where $f_r$ represents the frequency of the string $s$ and the denominator the cardinality of the union set of the equivalent strings under $T$.

For example, the string $s_1=0000$ for $n=4$ is grouped with the string $s_2=1111$ because they both have the same algorithmic complexity: $C_{0000}=\{0000, 1111\}$. The index of each class $C_i$ is the first string in the class according to arithmetical order. Thus the class \{0000, 1111\} is represented by $C_{0000}$. Another example of a class with two member strings is the one represented by $0011$ from the class $C_{0011}=\{0011, 1100\}$. By contrast, the string $0010 $ has other three strings of length 4 in the same class:  $C_{0010}=\{0100, 0010, 1101, 1011\}$. Other class with four members is the one represented by $0001$, the other three strings being $C_{0001} = \{0001, 0111, 1000, 1110\}$ because for any $s_i \in C_{0001}$ with $i<n$ the number of strings in $C_{0001}$, $T(s_i) = s_j$, i.e. by applying a transformation $T$ one can transform any string from any other in $C_{0001}$.

It is clear that $B$ induces a total order in $D_r$ from $D$ under the transformations $T$ preserving $K$ because if $s_1$, $s_2$ and $s_3$ are strings in $\{0,1\}^n$: $K(s_1) \leq K(s_2)$ and $K(s_2) \leq K(s_1)$ then $K(s_1) = K(b)$ so $s_1, s_2 $ are in the same complexity class $C_{s_1, s_2}$ (antisymmetry); If $K(s_1) \leq K(s_2)$ and $K(s_2) \leq K(c)$ then $K(s_1) \leq K(s_3)$ (transitivity) and either $K(s_1) \leq K(s_2)$ or $K(s_2) \leq K(s_1)$ (totality).

Hereafter  the $r$ in $D_r$ will simply be  denoted by  $D$, it being understood that it refers to $D_r$ after applying $B(D)$.

\subsection{Rank order correlation}

To figure out the degree of correlation between the probability frequency\cite{snedecor}, we followed a statistical method for rank comparisons. Spearman's rank correlation coefficient is a non-parametric measure of correlation, i.e. it makes no 
assumptions about the frequency distribution of the variables. Spearman's rank correlation coefficient is equivalent to the Pearson correlation on ranks. The Spearman coefficient has to do with measuring correspondence between two rankings for assessing the significance of this correspondence. The Spearman Rank Correlation Coefficient is:

\begin{center}
$\rho = 1 - (6 \Sigma d_i^2)/n(n^2 - 1)$
\end{center}

where $d_i$ is the difference between each rank of corresponding values of $x$ and $y$, and $n$ the number of pairs of values. \\

The Spearman coefficient is in the interval $[-1,1]$ where:

\begin{itemize}
\item If the agreement between the two rankings is perfect (i.e., the two rankings are the same) the coefficient has value 1.
\item If the disagreement between the two rankings is perfect (i.e., one ranking is the reverse of the other) the coefficient has value -1.
\item  For all other arrangements the value lies between -1 and 1, and increasing values (for the same number of elements) imply increasing agreement between the rankings.
\item If the rankings are completely independent, the coefficient has value 0.
\end{itemize}

\subsubsection{Level of significance}

The approach to testing whether an observed value of $\rho$ is significantly different from zero is to calculate the probability that it would be greater than or equal to the observed $\rho$, given the null hypothesis (that they are correlated by chance), by using a permutation test in order to conclude that the obtained value of $\rho$ is unlikely to occur by chance. 

The level of significance is determined by a permutation test\cite{good}, checking all permutations of ranks in the sample and counting the fraction for which the $\rho$ is more extreme than the $\rho$ found from the data. As the number of permutations grows proportional to $N!$, this is not practical even for small numbers. An asymptotically equivalent permutation test can be created when there are too many possible orderings of the data. For less than 9 elements we proceeded by a permutation test. For more than 9 elements the significance was calculated by Monte Carlo sampling, which takes a small (relative to the total number of permutations) random sample of the possible orderings, in our case the sample size was $10000$, big enough to guarantee the results given the number of elements.

The significance convention is that below $.5$, the correlation might be the product of chance and then it has to be rejected. If it is $0.05$, then there is enough confidence that the correlation has not occurred by chance and therefore it is said that the correlation is significant. If it is $0.01$ or below, then the correlation is said to be highly significant and very unlikely to be the product of chance since it would occur by chance less than 1 time in a hundred.

The significance tables generated and followed for the calculation of the significance of the Spearman correlation coefficients can be consulted in the following URL:

\noindent http://www.mathrix.org/experimentalAIT/spearmantables

\begin{figure}
\begin{center}
\includegraphics[width=1\textwidth]{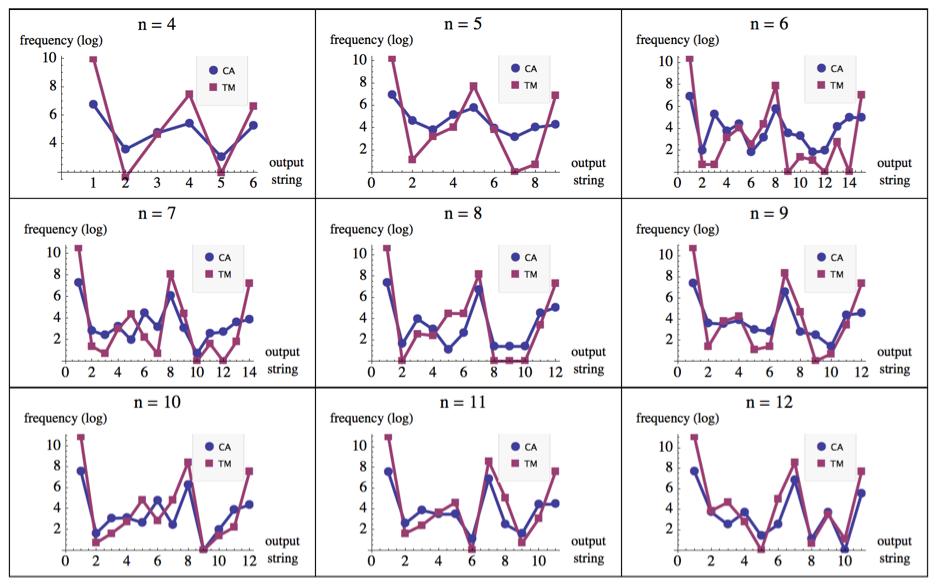}
\end{center}
\caption{The above sequence of plots show the evolution of the probability distributions for both 2-state Turing machines and one-dimensional elementary cellular automata, arranging the strings (x axis) in arithmetical order to compare the frequency value (y axis) of equal output strings produced by each $TM(2,2)$ and $CA(1)$. $n$ is the length of the strings to compare with, but also determines how far a machine runs in number of $t = 10 \times n$ steps and how many machines are sampled determined by: $a=n\times341$ for $TM(2,2)$ and $a = n \times 21$ for $CA(1)$ with $a$ the size of the sample so that $12 \times 341 = 4092$ and $a = 12 \times 21 =  252$ give the closest whole numbers to the total number of machines in $TM(2,2)$ and $CA(1)$ respectively. $n$ is in other words what let us define the progression of the sequence to look for the convergence in distribution. Our claim is that when $n$ tends to infinity the distributions converge either in order or in values to a limit distribution, as we will formulate in section \ref{convergence}.}
\label{comparison}
\end{figure}

\subsection{Convergence in distributions}
\label{convergence}

We want to find out if the probability distributions built from single and different models of computation converge.\\

\texttt{Definition (convergence in order)} A sequence of distributions $D_1, D_2, \dots$ converges to $D_N$, if for all string $s_i \in D_n$, $ord(s_i) \in D_n \rightarrow ord(s_i) \in D_N(s)$,  when $n$ tends to infinity. In other words, $D_n$ converges to an order when $n$ tends to infinity.\\

\texttt{Definition (convergence in values)} A sequence of distributions $D_1, D_2, \dots$ converges to $D_N$ if, for all string $s_i \in D_n$, $f(s_i) \in D_n \rightarrow f(s_i) \in D_N(s)$,  when $n$ tends to infinity.\\

\texttt{Definition (order-preserving):} A Turing machine $N$ is Kolmogorov-Chaitin complexity monotone, or Kolmogorov-Chaitin complexity order-preserving if, given the output probability distribution $D_1$ of $N$, if $K_{D_N}(s_1) \leq K_{D_N}(s_2)$ then $K_{D_2}(s_1) \leq K_{D_2}(s_2)$.\\

\texttt{Definition (quasi order preserving)} A Turing machine $N$ is $c$-Kolmogorov-Chaitin complexity monotone, or $c$-Kolmogorov-Chaitin complexity order- preserving if, for most strings, $N$ is Kolmogorov-Chaitin complexity monotone, or Kolmogorov-Chaitin complexity order-preserving. A Turing machine $N$ is $.01$-Kolmogorov-Chaitin complexity order-preserving is Kolmogorov-Chaitin complexity order-preserving.\\

In order to determine the degree of order-preserving we have introduced the term $c$ that will be determined by the correlation significance between two given output probability distributions $D_1$ and $D_2$.

In other words, one can still define a monotony measure even if only a significant first segment of the distributions converge. This is important because by algorithmic probability we know that random-looking strings will be--and because of their random nature have to be--very unstable exchanging places at the bottom of the distributions. But we may nevertheless want to know whether a distribution converges for most of the strings. \\

Whether or not a probability distribution $D$ converges to $D_N$, one might still want to check if two different models of computation converge between them:\\

\texttt{Definition (relative Kolmogorov-Chaitin monotony)} Let be $M$ and $N$ two Turing machine. $M$ and $N$ are relatively $c$-Kolmogorov-Chaitin complexity monotone if given their probability distributions $D_1$ and $D_2$ respectively and $K_{D_2}(s_1) \leq K{D_2}(s_2)$ then $K_{D_1}(s_1) \leq K_{D_1}(s_2)$ in $D_1$ for all $f(s_1), f(s_2) > c$.\\

\texttt{Definition (distribution length):} Given a model $M$, the length of its output probability distribution $D$ denoted by $|D|$ is the length of the largest string $s \in D$. \\

\texttt{Main result} $TM(2,2)$ and $CA(1)$ are relative Kolmogorov-Chaitin complexity quasi monotone up to $|D|=12$.\\

The following table shows the Spearman rank correlation coefficients for $D_{TM(2,2)}$ with $D_{CA(1)}$ from string lengths 2 to 12:\\

\begin{center}
$
\fbox{$
\begin{array}{c|c|c}
$Number$ & $Spearman$ & $Significance$\\
$of$ \text{ } $elements$ & $coefficient$ & $value$ \\
 2 & 1 & 50 \\
 3 & 1 & 33.33\\
 6 & 0.94 & 0.01\\
 9 & 0.78 & 0.01\\
 15 & 0.44 & 0.01 \\
 14 & 0.66 & 0.01\\
 12 & 0.67 & 0.01\\
 12 & 0.78 & 0.01\\
 12 & 0.80 & 0.02\\
 11 & 0.79 & 0.01\\
 11 & 0.80 & 0.01
\end{array}
$}
$
\end{center}

Significance values are not expected to score well at the beginning due to the lack of elements to determine if other than the product of chance produced the order. For 2 elements in each rank order there are only 2 ways to arrange each rank, and even if they make a perfect match as they do, the significance cannot be higher than 50 percent because there is still half chance to have had produced that particular order. It is also the case for 3 elements, even when the ranks made a perfect match as well. But starting at 6 one can start looking to an actual significance value, and up to 12 in the sequence below one can witness a notorious increase up to stabilize the value at $0.01$ which is, for all them, highly significant. Just one case was just \emph{significant} rather than \emph{highly significant} according to the threshold convention.

\begin{figure}
\begin{center}
\includegraphics[width=.7\textwidth]{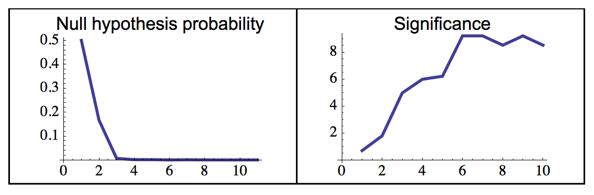}
\end{center}
\caption{The probability of the null hypothesis (that between $D_{TM(2,2)}$ and $D_{CA(1)}$ the correlation is the product of chance) decreases very soon remaining very low, while the significance increases systematically from $n = 2$ to $12$.}
\label{dist}
\end{figure}

The fact that each of the values of the sequence are either significant or highly significant makes the entire sequence convergence even more significant. $D_{TM(2,2)}$ and $D_{CA(1)}$ are therefore statistically highly correlated and they are relative 0.01-Kolmogorov-Chaitin complexity quasi monotone up to $|D|=12$ in almost all values. Therefore $TM(2,2)$ and $CA(1)$ are relative Kolmogorov-Chaitin complexity monotone.

It also turned out that the Pearson correlation coefficients were all highly significant between the actual probability values between $D_{TM(2,2)}$ and $D_{C(1)}$, with the following values:

\begin{center}
$
\fbox{$
\begin{array}{c|c}
$Number$ & $Pearson$ \\
$of$ \text{ } $elements$ & $coefficient$ \\
 2 & 1 \\
 3 & 0.624662 \\
 6 & 0.979218 \\
 9 & 0.972992 \\
 15 & 0.95721 \\
 14 & 0.975683 \\
 12 & 0.920039 \\
 12 & 0.942916 \\
 12 & 0.982229 \\
 11 & 0.916871 \\
 11 & 0.944149
\end{array}
$}
$
\end{center}

\noindent The above results are important because they not only show that $TM(2,2)$ and $CA(1)$ are Kolmogorov-Chaitin monotone up to $|D|=12$ but because they constitute the basis and evidence for the formulation of the conjectures in section \ref{conjectures}: \\

\subsection{Conjectures of convergence}
\label{conjectures}

Let $ord$ denote the ranking order of a distribution $D$ and $pr$ the actual probability values of $D$ for each string $s \in D$, then:\\

\texttt{Conjecture 1}  If $pr(D_{TM}(n))=\{f(s_1),f(s_2),\dots,f(s_u)\}$, then for all $s_i$, $f_{s_i}$ $\rightarrow$ $f(L_{s_i})$ when $n$ $\rightarrow$ $\infty$ with $\{f(L_{s_1)},f(L_{s_2}),$ $\ldots,f(L_{s_n}),\ldots \}$ the limit frequencies. In other words, the sequence of probability values $f(D_{TM}(1)), f(D_{TM}(2)),$ $\ldots, f(D_{TM}(n)), \ldots$ converges when $n$ tends to infinity. Let's call this limit distribution $pr$ hereafter.\\

\texttt{Conjecture 2}  The sequence $ord(D_{TM}(1)), ord(D_{TM}(2)),\ldots,ord(D_{TM}(n))$ converges when $n$ tends to infinity.\\

\noindent Notice that the conjecture 2 is weaker than the conjecture 1 since conjecture 2 could be true even if conjecture 1 is false. Both conjectures 1 and 2 imply there exists a $k \in \{1, 2, \dots, n\}$ such that for all $i > k$, $TM(i)$ is Kolmogorov-Chaitin complexity order-preserving.\\

\noindent Likewise for cellular automata:\\

\texttt{Conjecture 3}  The sequence $pr(D_{CA}(1)), pr(D_{CA}(2), \ldots, pr(D_{CA}(u))$ converges to $pr(D_{CA}(n))$ when $n$ tends to infinity.\\

\texttt{Conjecture 4}  The sequence $ord(D_{CA}(1)), ord(D_{CA}(2)),\ldots,ord(D_{CA}(n))$ converges when $n$ tends to infinity.\\

\noindent Notice that the conjecture 2 is weaker than the conjecture 1 since conjecture 2 could be true even if conjecture 1 is false. Both conjectures 1 and 2 imply there exists a $k \in \{1, 2, \dots, n\}$ such that for all $i > k$, $CA(i)$ is Kolmogorov-Chaitin complexity order-preserving.\\

Likewise for Turing machines, conjecture 3 implies conjecture 4, but conjecture 4 could be true even if conjecture 3 is false. \\

\texttt{Conjecture 5} $pr(D_{CA}(n)) = pr(D_{T}(n))$.\\

\texttt{Conjecture 6} $ord(D_{CA}(n)) = ord(D_T(n))$.\\

In other words, the limit distributions for both $CA$ and $TM$ converge to the same limit distributions.\\

Conjecture 5 implies conjecture 6, but conjecture 6 could be true even if conjecture 5 is false. \\
 
\noindent Both $pr$ and $ord$ define $D_N$, from now on the natural probability distribution. We now can propose our definition of a natural model of computation:\\

\texttt{Definition (naturalness in distribution)} $M$ is a natural model of computation if it is $c$-Kolmogorov-Chaitin monotone or $c$-Kolmogorov-Chaitin order-preserving for $c=.01$.\\

\begin{figure}
\begin{center}
\includegraphics[width=.5\textwidth]{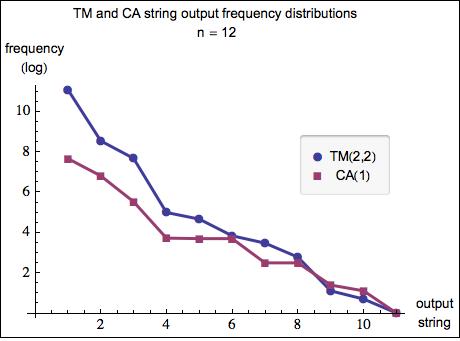}
\end{center}
\caption{Frequency (log) distributions for $TM(2,2)$ and $CA(1)$, for $n=12$. In this plot no string arrangement was made, unlike figure \ref{comparison}. The rate of grow seems to follow a power law.}
\label{dist}
\end{figure}

In other words, any model of computation preserving the relative order of the natural distribution $D_N$ is natural in terms of Kolmogorov-Chaitin complexity under our definition. So one can now technically say that a tailor-made Turing machine producing a different enough output distribution is not natural according to the prior $D_N$. One can now also define (a) a degree of $naturalness$ according to the ranking coefficient and number of order-preserving strings as suggested before and (b) a Kolmogorov-Chaitin order-preserving test such that one can be able to say whether a programming language or Turing machine is natural by designing an experiment and running the test. For (a) it suffices to follow the ideas in this paper. For (b) one can follow the experiments described partially here supplemented  with further details available in \cite{delahayezenil2} in order to produce a probability distribution that could be compared  to the natural probability distribution to determine whether or not convergence occurs.
The use of these natural distributions as prior probability distributions are one of the possible applications. The following URL provides the full tables: http://www.mathrix.org/experimentalAIT/naturaldistribution\\
Further details, including the original programs, are available online in the \emph{experimental Algorithmic Information Theory} web page: http://www.mathrix.org/experimentalAIT/\\

Further experiments are in the process of being performed, both for bigger classes of the same models of computation and for other models of computation, including some that clearly are not Kolmogorov-Chaitin order-preserving. More experiments will be performed covering different parameterizations, such as distributions for non-empty initial configurations, possible rates of convergence and radius of convergence, as well as the actual relation between the mathematical expected values of the theoretical definitions of $K(s)$ and $m(s)$ (the so called universal distribution\cite{li}), as first suggested in \cite{delahayezenil1, delahayezenil2}. We are aware of the possible expected differences between probability distributions produced by self-nondelimiting vs. self-delimiting programs\cite{chaitin1}, such as in the case discussed within this paper, where the halting state of the Turing machines was partially dismissed while the halting of the cellular automata was randomly chosen to produce the desired length of strings for comparison with the TM distributions. A further investigation suggests the possibility that there are interesting qualitative differences in the probability distributions they produce. These can be also be studied using this approach. 

If these conjectures are true, as suggested by our experiments, this procedure is a feasible and effective approach to both $m(s)$ and $k(s)$. Moreover, as suggested in\cite{delahayezenil1}, it is a way to approach the Kolmogorov-Chaitin complexity of short strings. Furthermore, statistical approaches might  in general be good approaches to the Kolmogorov-Chaitin complexity of strings of any length, as long as the sample is large enough for getting a reasonable significance.\\

\bibliographystyle{fundam}
\bibliography{citations}

\end{document}